\documentclass[prl,reprint,english,twocolumn,showpacs,superscriptaddress]{revtex4-2}
\usepackage{bm,mathrsfs}
\usepackage[dvipdfmx]{graphicx}
\usepackage{epsfig}
\usepackage{psfrag}
\usepackage{amsmath,bbm}
\usepackage{amsfonts,amssymb}
\usepackage[utf8]{inputenc}
\usepackage[T1]{fontenc}
\usepackage{times}
\usepackage{dsfont}
\usepackage{enumitem}  
\usepackage{comment}
\usepackage[dvipdfmx,colorlinks=true,linkcolor=blue,urlcolor=blue,citecolor=blue]{hyperref}
\usepackage{braket,physics}
\usepackage{verbatim,color}
\newcommand{\beq}{\begin{equation}}
\newcommand{\eeq}{\end{equation}}
\newcommand{\bseq}{\begin{subequations}}
\newcommand{\eseq}{\end{subequations}}
\newcommand{\bal}{\begin{aligned}[b]}
\newcommand{\eal}{\end{aligned}}
\newcommand{\beqa}{\begin{eqnarray}}
\newcommand{\eeqa}{\end{eqnarray}}

\begin{document}

\title{Non-ergodic dynamical phase transition via a zero-mode exceptional point \\
in a non-Markov atomic Josephson junction}

\author{Koichiro Furutani}
\email{furutani.koichiro.s7@f.mail.nagoya-u.ac.jp}
\affiliation{Department of Applied Physics, Nagoya University, Nagoya 464-8603, Japan}
\affiliation{Institute for Advanced Research, Nagoya University, Nagoya 464-8601, Japan}

\date{\today}

\begin{abstract}
Open quantum systems typically lose their initial memory due to the environmental decoherence resulting in thermalization. 
We demonstrate a striking breakdown of this paradigm in a head-to-tail Bose-Josephson junction, which is described by an intrinsically momentum-coupled Caldeira-Leggett model. 
Through exact non-Markov Langevin simulations, we discover a novel type of non-ergodic dynamical phase transitions into a running state, which has no counterpart in Markov limit. 
Crucially, we reveal that this transition is fundamentally governed by a zero-mode exceptional point emerging from the non-Markov friction. 
This topological origin is characterized by the winding of the response function. 
Finally, numerical quantum simulations of an equivalent driven XXZ spin chain confirm that this exceptional-point-induced signature robustly survives as a dynamical crossover against strong quantum fluctuations and the dynamical backreaction of the environment. 
This macroscopic robustness offers a promising platform for long-lived quantum memories in dissipative environments.

\end{abstract}

\maketitle

%{\it--- Introduction}
The interplay between macroscopic quantum coherence and environmental dissipation is a central issue in modern physics. 
A paradigm for this interplay is a quantum particle in a spatially periodic potential subject to friction. 
In the conventional Caldeira-Leggett (CL) model with coordinate-coupled Ohmic dissipation realized with a resistively-shunted superconducting Josephson circuit \cite{koch1980, koch1982,caldeira1983, furutani2021}, the system undergoes the Schmid-Bulgadaev (SB) quantum phase transition: a strict localization when dissipation strength exceeds a critical threshold \cite{schmid1983,bulgadaev1984,muramatsu1985,fisher1985,kane1992, furusaki1993, fendley1995, fateev1997,werner2005,lukyanov2007,altlandsimons, nagaosa,murani2020,masuki2022,yokota2023,daviet2023,sepulcre2023,masuki2023,houzet2024, kuzmin2025, paris2025}.

Recently, the breakdown of thermalization --- allowing systems to robustly retain initial memory --- in quantum systems has attracted immense interest \cite{abanin2019, serbyn2021, moudgalya2022, adler2024, tufarelli2013,gonzalez2013, facchi2016, calajo2019, dinc2019, bark2021}. 
Ultracold atoms, particularly Bose-Josephson junctions (BJJs), offer pristine platforms to simulate such dissipative dynamics \cite{pigneur2018, scherg2021, furutanitempere2022, amico2022, su2023, kohlert2023, valencia2024, adler2024}. 
Crucially, microscopic derivations for a head-to-tail BJJ revealed that the Josephson mode inherently couples to the Bogoliubov phonon bath via {\it momentum}, rather than phase coordinate \cite{polo2018,binanti2021}. 
The static thermodynamics of this intrinsically momentum-coupled CL (ICL) model maps exactly onto the boundary sine-Gordon (bsG) model, exhibiting an equilibrium ground-state phase diagram identical to the SB phase diagram \cite{ashida2021, ashida2022, masuki2023, furutani2024}.

%%%% Phase diagram (\alpha, v_{0,c}) %%%%%%%%
\begin{figure}[t]
\centering
\includegraphics[keepaspectratio,width=70mm]{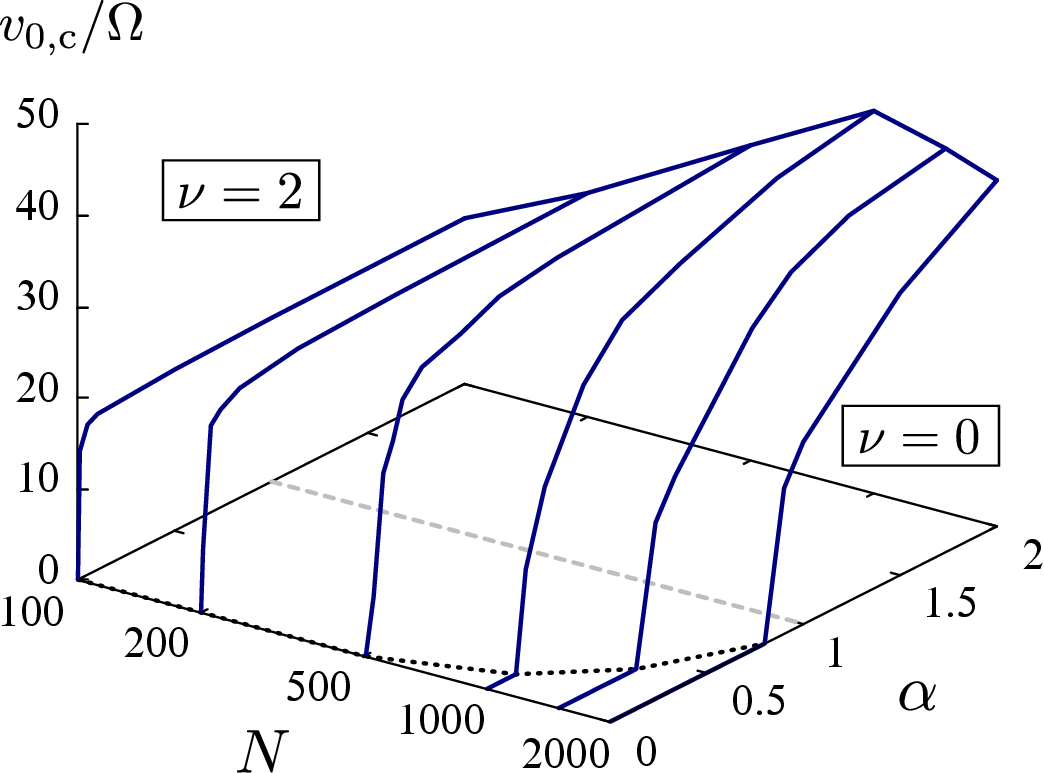}
\caption{Critical initial velocity scaled by the Josephson plasma frequency $\Omega$ as a function of the damping parameter $\alpha$ and the number of bath modes $N$ under $\kappa L=100\pi$. 
The gray dashed line represents the SB critical line of the ground state. 
The black dotted curve represents the parameter at which the critical velocity starts to deviates from zero. 
The Josephson phase oscillates in the localized phase with a small initial velocity $v_{0}<v_{0,\mathrm{c}}$, whereas it turns into an incoherent running state characterized by the winding number $\nu=2$ with a larger initial velocity $v_{0}>v_{0,\mathrm{c}}$. }
\label{Figtcritical}
\end{figure}
%%%%%%%%%%%%%%%%%%%%%%%%%%%%%%%%%%%%%%%%%%%%%

However, the assumption that the real-time nonequilibrium dynamics will universally thermalize into this SB ground state fundamentally relies on the ergodic hypothesis. 
While generic open quantum systems lose their initial memory to the environment and relax to the equilibrium state predicted by the static phase diagram, the ICL model challenges this conventional assumption. 
The interplay of momentum coupling and the Josephson potential inevitably generates a highly nonlinear, non-Markov friction and super-Ohmic noise \cite{binanti2021}. 
This nontrivial dissipation structure raises a profound question: does the macroscopic phase dynamics truly relax to the SB ground state, or does the initial memory survive long enough to dictate a completely different nonequilibrium fate? 

In this Letter, we answer this question by revealing a striking breakdown of ergodicity and a novel dynamical phase transition in the head-to-tail BJJ. 
Solving the exact nonlinear non-Markov Langevin dynamics, we demonstrate that the system completely evades the SB phase, provided that the initial phase velocity or population imbalance exceeds a critical threshold. 
Remarkably, above the threshold, the system undergoes a sharp dynamical transition into an incoherent running state, as illustrated in the dynamical phase diagram of Fig.~\ref{Figtcritical}, instead of thermalization. 
Crucially, we uncover that this transition is fundamentally driven by a zero-mode exceptional point (EP) at zero-energy emerging from the non-Markov friction. 
At this EP, merging of the two poles of the response function in the complex frequency plane generates a zero-energy double pole, establishing a topological constraint characterized by the winding of the response function, in contrast to the critical damping where poles merge at a finite decay rate in a damped harmonic oscillator. 
Finally, time-evolving block decimation (TEBD) simulations of an equivalent XXZ spin chain under a time-dependent transverse magnetic field at the boundary confirm that while strong quantum fluctuations round the sharp transition into a dynamical crossover, the topological signature of the EP robustly governs the macroscopic dynamics \cite{vidal2003, vidal2004, vidal2007, banuls2019}. 
Crucially, this robustness persists even in the presence of the dynamical backreaction from the out-of-equilibrium bath modes, which goes beyond the Langevin analysis based on the fluctuation-dissipation relation. 
Our findings predict another type of dynamical phase transition driven by non-Markov EP topologically distinct from the SB transition of the ground state and establish the BJJ as a macroscopic quantum amplifier, where non-Markov topology dictates long-lived quantum memories.

{\it Microscopic model and generalized Langevin equation.---}
We start with a quasi-one-dimensional Bose Josephson junction described by
\beq
\bal
H&=\sum_{a=1,2}\int_{0}^{L}\dd{x}\qty[\dfrac{\hbar^{2}}{2m}\abs{\partial_{x}\psi_{a}}^{2}+\dfrac{g}{2}\abs{\psi_{a}}^{4}] \\
&-\int_{0}^{L}\dd{x}\dfrac{J(x)}{2}\qty(\psi_{1}^{*}\psi_{2}+\psi_{2}^{*}\psi_{1}) ,
\label{HBJJ}
\eal
\eeq
with $\psi_{a=1,2}=\sqrt{n_{a}} e^{i\phi_{a}}$ being the Bose field in each tube with length $L$, $m$ being the atomic mass, $g>0$ being the contact interaction strength, and $J(x)$ being a Josephson coupling. 
We consider the head-to-tail configuration of the Josephson coupling $J(x)=J_{0}L\delta(x)$ and introduce the uniform average density $\bar{n}=n_{1}+n_{2}$ and the relative phase $\phi(x,t)=\phi_{1}(x,t)-\phi_{2}(x,t)$. 
We derived the equation of motion, the generalized Langevin equation for the Josephson mode $\phi_{0}(t)\equiv \phi(x=0,t)$, as \cite{polo2018,binanti2021, furutani2023,furutani2024}
\beq
\bal 
\dfrac{\hbar^{2}}{2E_{C}} &\ddot{\phi}_{0}(t)+\int_{0}^{t}\dd{s}\gamma(t-s)\dv{s}\sin{\phi_{0}(s)} \\
&+\dfrac{E_{J}}{1+N}\sin{\phi_{0}(t)}
=\xi(t) ,
\eal
\label{langevin}
\eeq
with $E_{C}=(1+N)\hbar^{2}/2ML^{2}$ and $E_{J}=J_{0}L\bar{n}$ being the charging and the Josephson energies, respectively. 
The effective mass $M=\hbar^{2}/2gL$ is related to the interatomic interaction strength, and the damping kernel 
\beq
\gamma(t) 
=\dfrac{4E_{J}E_{C}}{\hbar^{2}} \int_{0}^{\infty}\dd{\omega}\dfrac{J(\omega)}{\omega^{3}}\cos{(\omega t)},
\label{dampingkernel}
\eeq
is non-Markov for a finite number of bath modes $N$ with the super-Ohmic spectral density
\beq
J(\omega)=\dfrac{\hbar^{2}}{4E_{C}(1+N)}\sum_{n=1}^{N}\omega_{n}^{3}\delta(\omega-\omega_{n}).
\eeq 
The bath modes have the phonon dispersion $\omega_{n}=c k_{n}$ with $k_{n}=n\pi/L$ ($n=1,2,\cdots,N$) and $c=\sqrt{g\bar{n}/m}$ being the sound velocity. 
In $N\to\infty$, the damping kernel \eqref{dampingkernel} reduces to $h E_{J}/(\alpha E_{C}) \delta(t)$ corresponding to the Markov limit, where $\alpha\equiv 2\pi/\sqrt{\tilde{g}}$ is the damping parameter and $\tilde{g}=mg/\bar{n}\hbar^{2}$ is the gas parameter. 
The Gaussian noise $\xi(t)$ satisfies
\bseq
\beq
\langle \xi(t)\rangle=0, 
\eeq
\beq
\bal
\langle \xi(t) \xi(0)\rangle 
=\hbar\int_{0}^{\infty}\dd{\omega} J(\omega)\coth{\qty(\dfrac{\hbar\omega}{2 k_{\rm B}T})}\cos{(\omega t)} ,
\label{noisecorrelator}
\eal
\eeq
\label{noiserelations}
\eseq
with $k_{\rm B}$ being the Boltzmann constant. 
The super-Ohmic spectral density $J(\omega)\propto \omega^3$ strongly amplifies quantum fluctuations as in Eq.~\eqref{noisecorrelator}. 
The initial slippage term is negligible as long as the initial phase is set to zero in the following analyses. 
We revealed that the ground state of the head-to-tail BJJ exhibits the SB transition at $\alpha = 1$ in $N\to\infty$ as a consequence of the giant phase fluctuations due to super-Ohmic dissipation \cite{furutani2024}. 
It has been verified by the time evolution of the Langevin equation \eqref{langevin} in the continuum Markov limit. 
With a finite number of modes, on the other hand, the system is governed by non-Markov stochastic dynamics, which is our target in this Letter.

\begin{figure}[t]
\centering
\begin{minipage}{0.49\columnwidth}
\includegraphics[keepaspectratio,width=46.5mm]{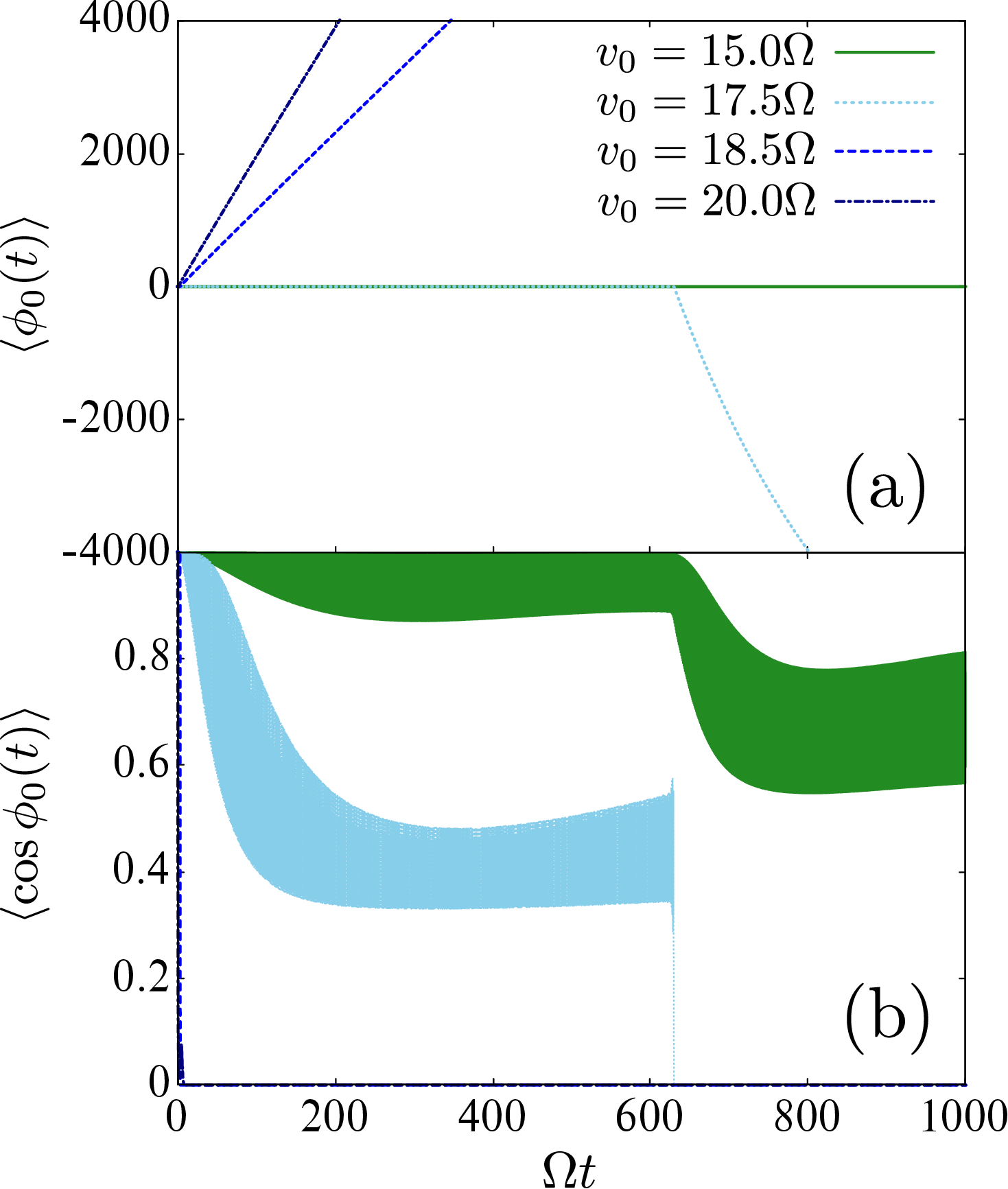}
\end{minipage}
\begin{minipage}{0.49\columnwidth}
\includegraphics[keepaspectratio,width=39mm]{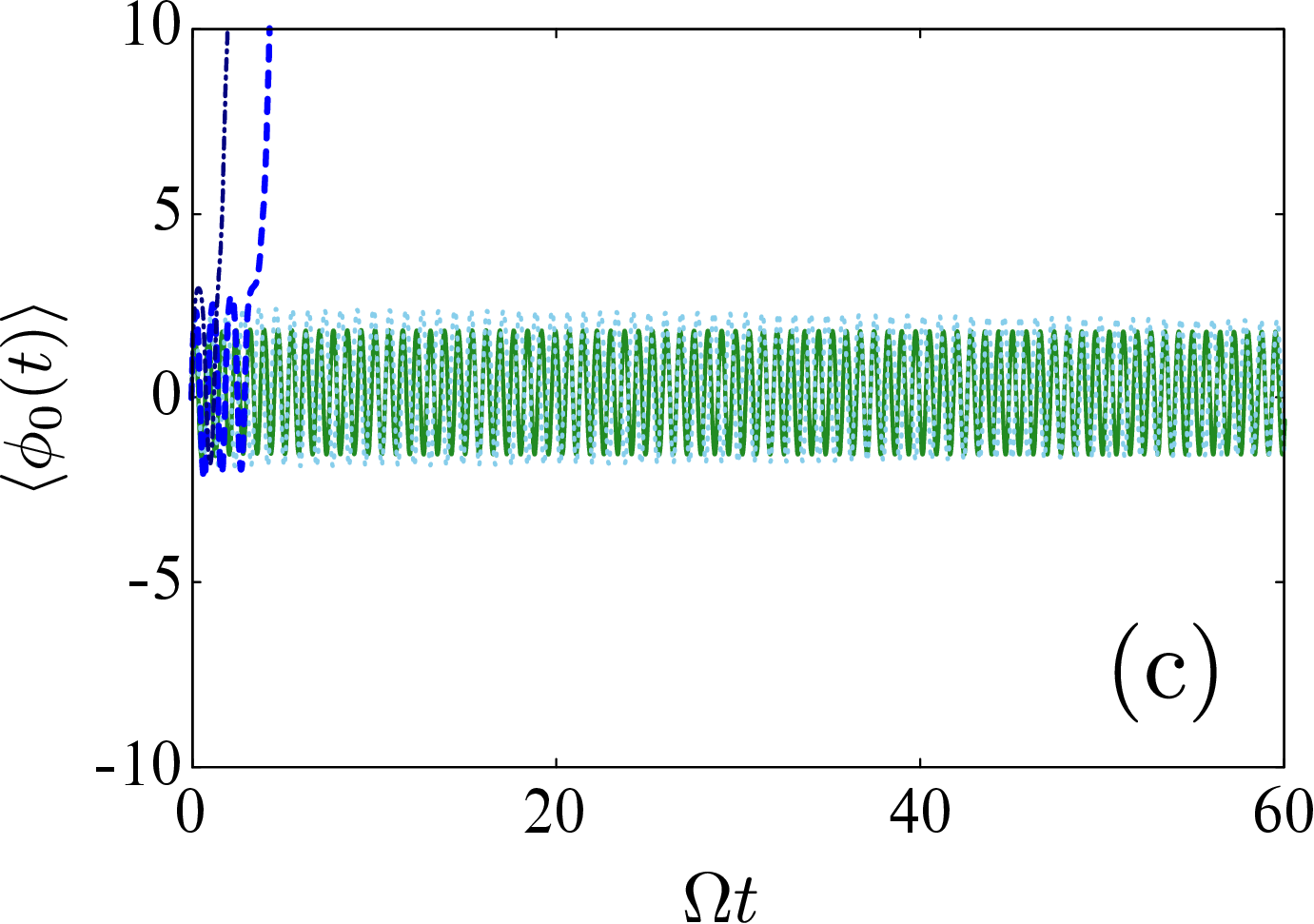}
\includegraphics[keepaspectratio,width=39mm]{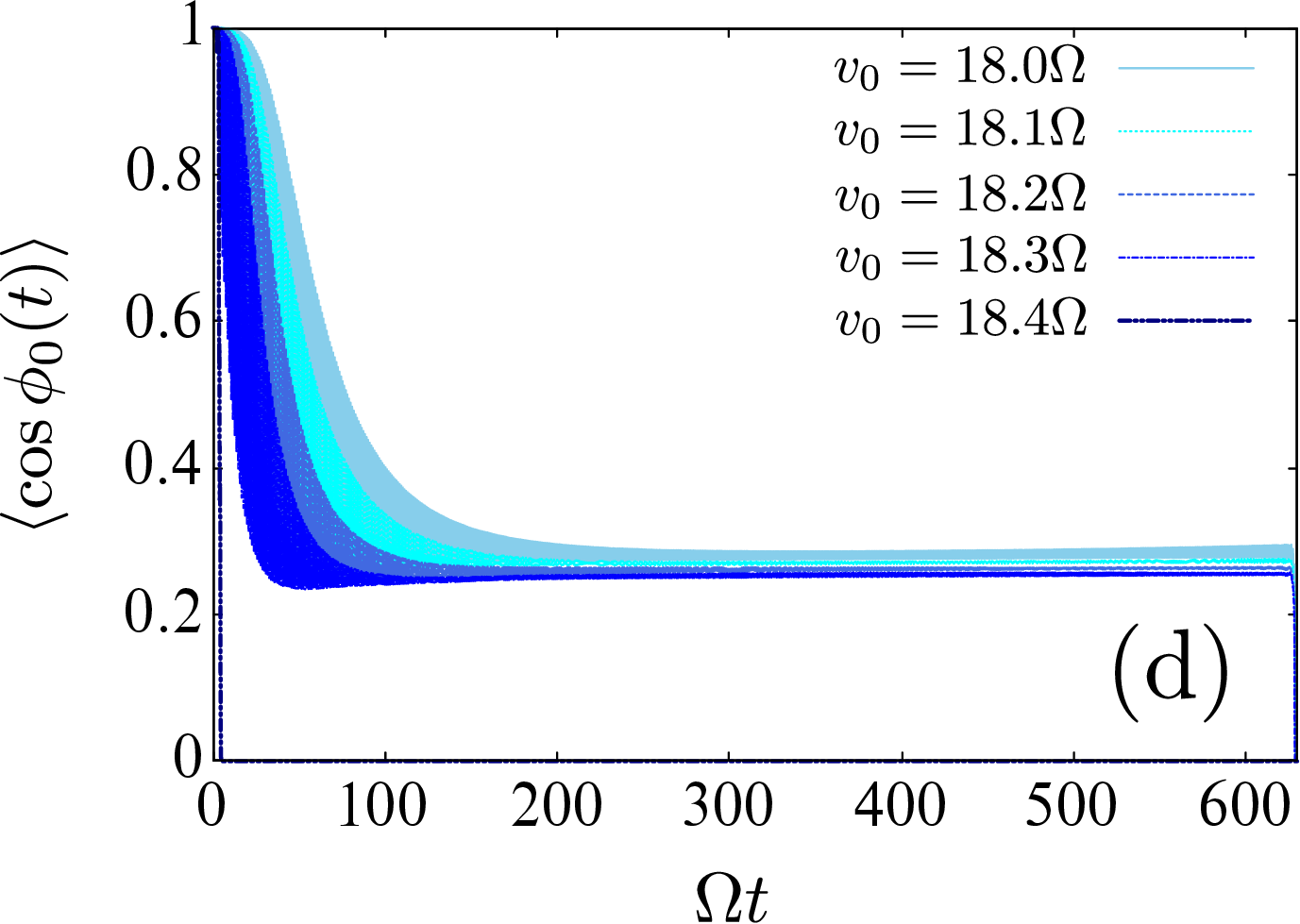}
\end{minipage}
\caption{(a) Time evolution of $\langle \phi_{0}(t)\rangle$ under $\alpha=10$ and $\phi_{0}(0)=0$ with several choices of the initial phase velocity $v_{0}$. 
We set the number of bath modes $N=100$ and $\kappa L=100\pi$, which corresponds to the ultraviolet cutoff frequency $W=\Omega$. 
(b) Time evolution of the coherence factor $\langle \cos{\phi_{0}(t)}\rangle$. 
(c) $\langle \phi_{0}(t)\rangle$ in the short time $\Omega t\le 60$.
(d) Coherence factor in the vicinity of the critical velocity $v_{0,\mathrm{c}}\simeq 18.4\Omega$. }
\label{Figphiav}
\end{figure}

{\it Exact non-Markov dynamics and running phase transition.---}
To simulate non-Markov dissipative dynamics, we numerically solve Eq.~\eqref{langevin} via the Markov embedding method mapping the non-Markov system into a Markov system with additional dynamical variables \cite{garraway1997, strasberg2016, tamascalli2018, nazir2018, brenes2020, kanazawa2020, klippenstein2021, trivedi2021, kanazawa2024, dalton2025, SM}. 
Figure \ref{Figphiav} shows the time evolution of $\langle\phi_{0}(t)\rangle$, which is computed by the ensemble average of $\mathcal{N}=10^{4}$ noise realizations, with the time scaled by the Josephson plasma frequency $\Omega = \sqrt{2E_{J}E_{C}/(1+N)}/\hbar$. 
Panel (a) shows the time evolution under $\alpha=10$ (deep coherent regime), the number of bath modes $N=100$, and $\kappa L=100\pi$ where $\kappa = \sqrt{2mJ_{0}}/\hbar$, which determines the level spacing of the bath modes, yields the ultraviolet cutoff frequency $W\equiv \omega_{N}=\Omega$, with several choices of initial phase velocity $v_{0}\equiv \dot{\phi}_{0}(0)$. 
With a small initial phase velocity $v_0\lesssim 17\Omega$, the phase exhibits a Josephson oscillation as also shown in the panel (c), but it turns into a running state above a threshold $v_{0,\mathrm{rec}}\simeq 17.5\Omega$ at $t=t_{\rm rec}\equiv 2\pi/\omega_{1}$, which is the recurrence time of the phonon bath. 
This dynamic transition in the long time regime accompanies the Josephson phase running in the opposite direction of the initial phase velocity, which stems from the collective frictional force of the synchronized bath modes at the recurrence time. 
The recurrence results from the discrete levels of the phonon bath and the continuum limit $\kappa L\to\infty$ makes the recurrence time infinitely long $t_{\rm rec}\to\infty$. 
To rigorously isolate the intrinsic non-Markov phase dynamics from the finite-size bath recurrence, we strictly restrict our analysis to the pre-recurrence regime $t<t_{\rm rec}$. 
By further increasing the initial velocity, it starts to run at $t\simeq 0$ in the direction of the initial velocity as plotted by the dashed line of $v_0 = 18.5\Omega$ in panel (a). 
The origin of the transition is distinct from the recurrence and peculiar to the non-Markov dissipation. 
This running phase transition indicates the incoherent state characterized by the vanishing coherence factor $\langle \cos{\phi_0(t)}\rangle$ plotted in panel (b). 
Figure \ref{Figphiav}(d) shows the coherence factor with initial velocities close to the critical initial velocity $v_{0,\mathrm{c}}\simeq 18.4\Omega$. 
For $v_{0}<v_{0,\mathrm{c}}$, the coherence factor approaches around $0.25$ by increasing $v_0$ in the long time regime before recurrence. 
At critical velocity, coherence rapidly vanishes and never becomes finite. 
It clearly shows that the running state never relaxes to the coherent superconducting ground state, which demonstrates ergodicity-breaking. 
The $\alpha$-dependence of the critical velocity is summarized in Fig.~\ref{Figtcritical}. 
The critical velocity increases monotonically with respect to the damping parameter $\alpha$. 
The black dotted curve on the $N$-$\alpha$ plane represents the damping parameter $\alpha_{\rm c}(N)$ at which the critical velocity deviates from zero as a function of the number of bath modes $N$. 
By increasing $N$, the critical damping parameter $\alpha_{\rm c}$ approaches unity, which corresponds to the Markov limit making the Josephson mode thermalize to the ground state classified by the SB phase diagram \cite{furutani2024}. 
These results are restricted to zero temperature, but we confirmed the running phase transition also at finite temperature. 
Although thermal fluctuations reduce the critical velocity compared to the zero temperature case, they do not qualitatively affect the presence of the dynamical phase transition \cite{SM}.

\begin{figure}[t]
\centering
\includegraphics[keepaspectratio,width=70mm]{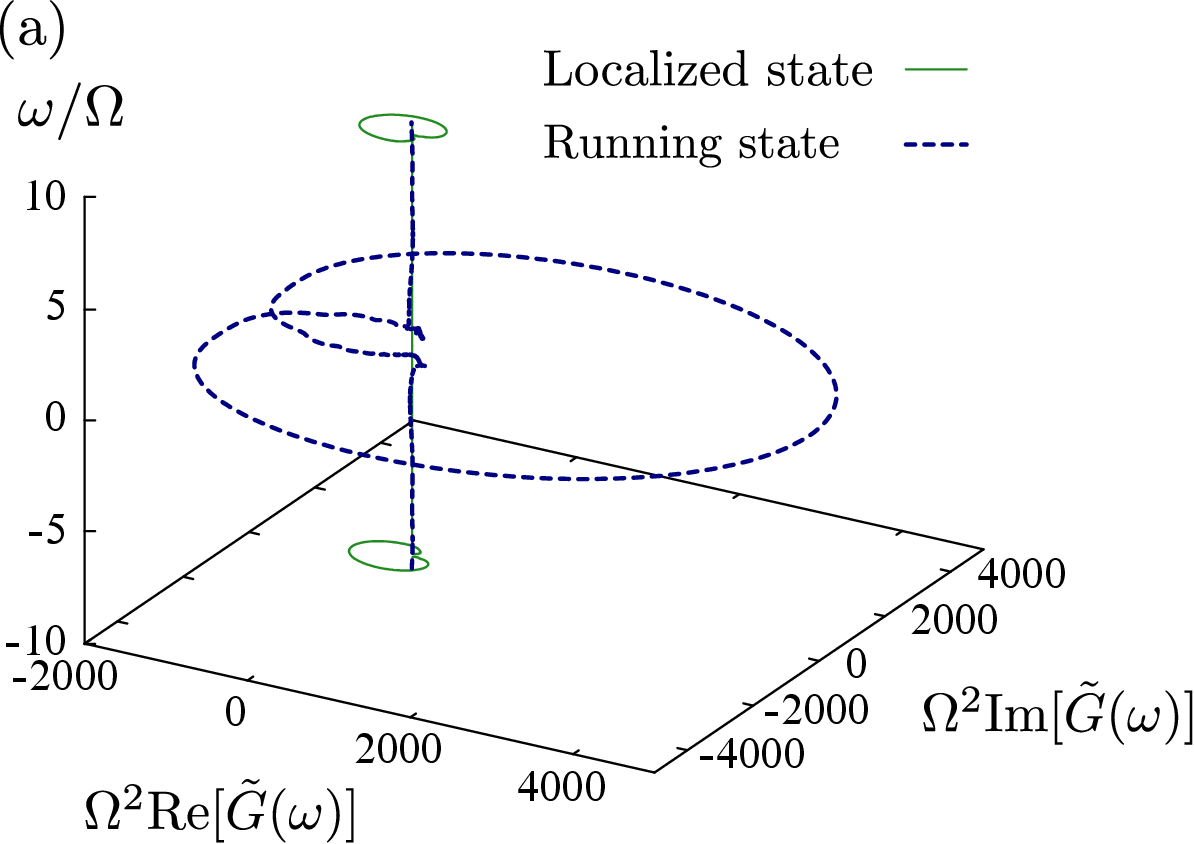}
\begin{minipage}{0.49\columnwidth}
\includegraphics[keepaspectratio,width=43mm]{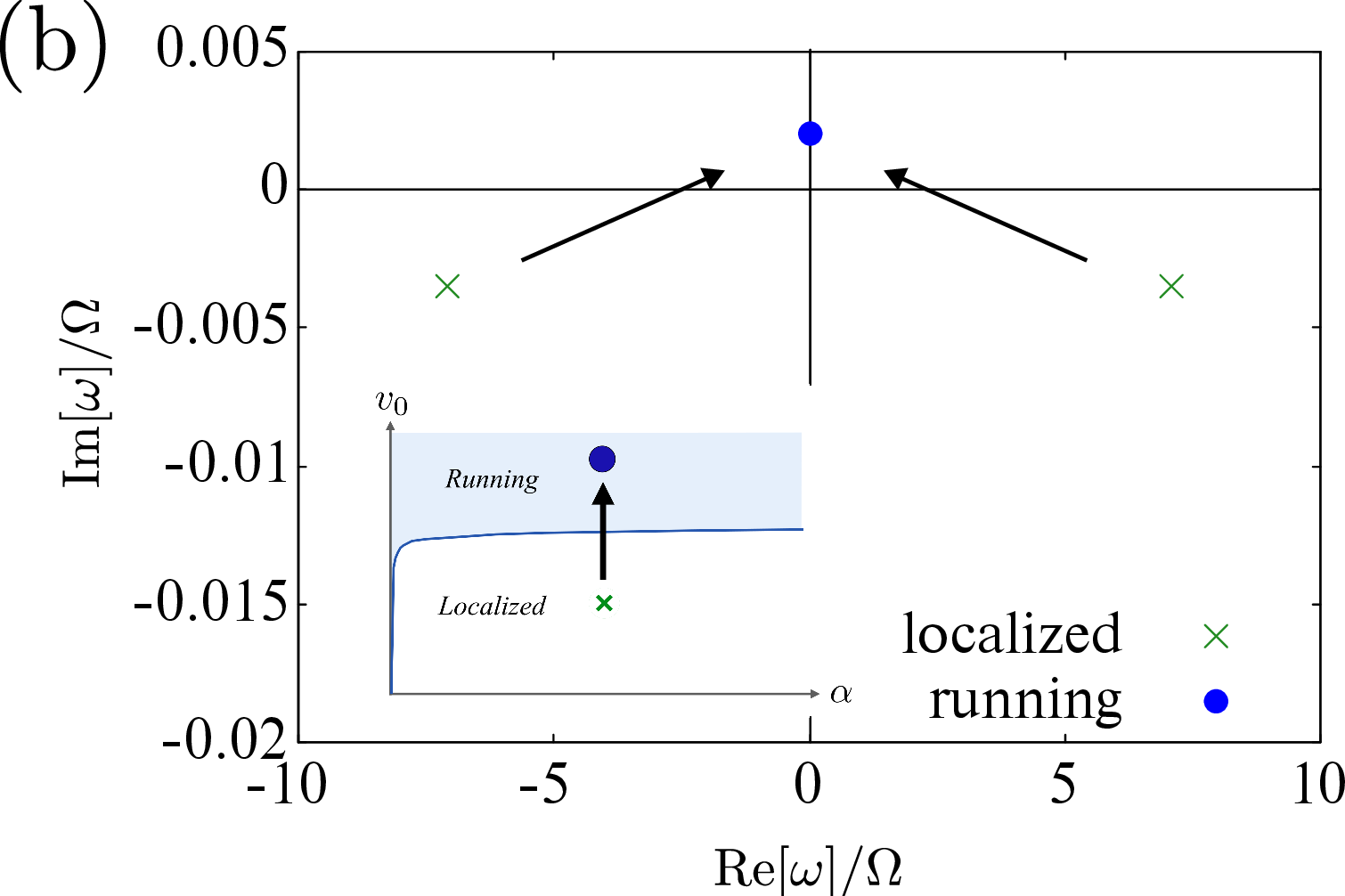}
\end{minipage}
\begin{minipage}{0.49\columnwidth}
\includegraphics[keepaspectratio,width=43mm]{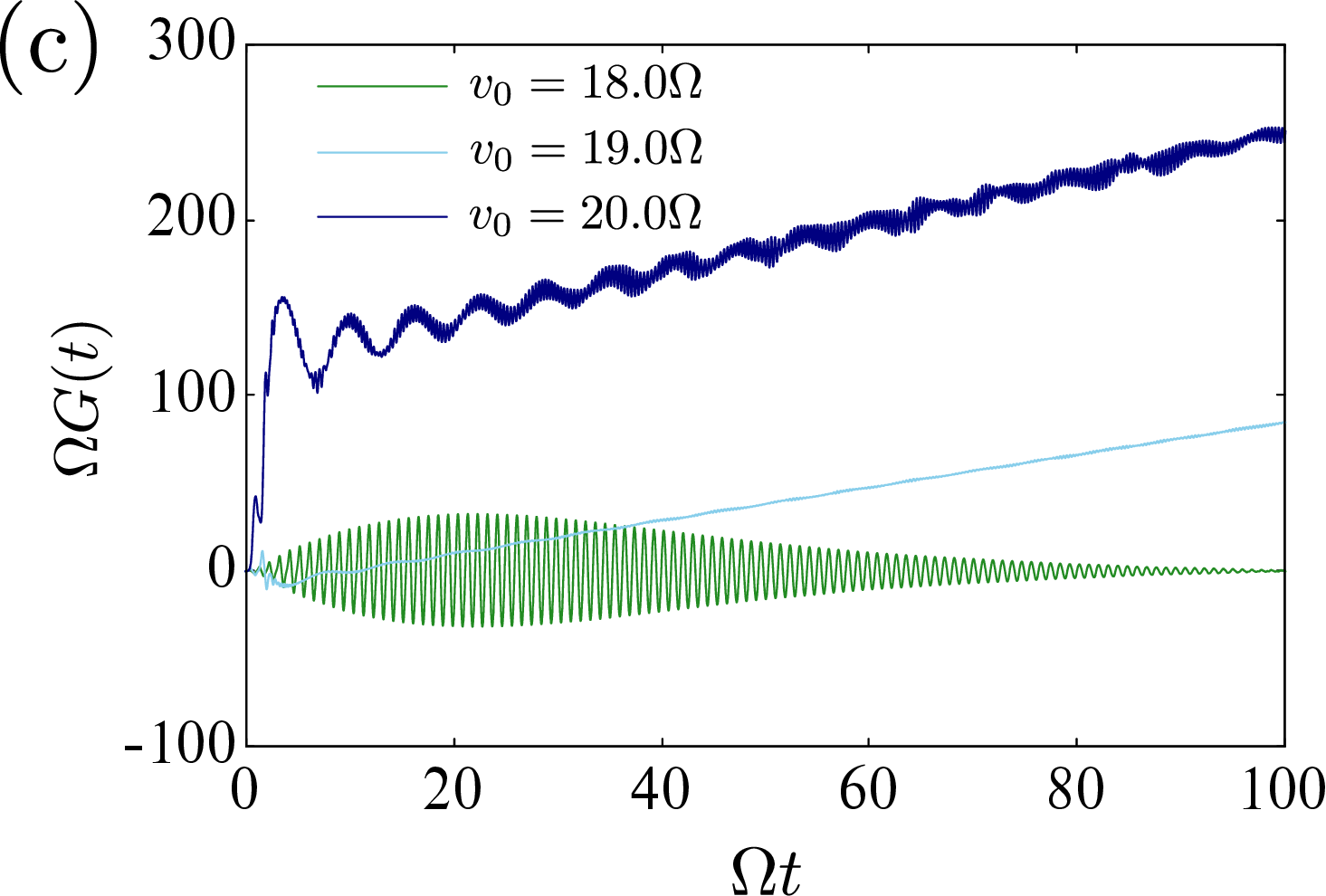}
\end{minipage}
\caption{(a) Response function in the real frequency space for $\alpha=2$ with $v_{0}=10\Omega$ (localized state) and $v_{0}=20\Omega$ (running state), respectively, under $N=100$ and $\kappa L=100\pi$. 
(b) Poles of the response function in the complex frequency space. 
Inset: Critical velocity as a function of $\alpha$. 
(c) Real-time response function with $v_{0}=18\Omega$ (localized state) and $v_{0}=19\Omega,\,20\Omega$ (running state), respectively. }
\label{Fignyquist}
\end{figure}

{\it Zero-mode exceptional point.---}
This dynamic running phase transition triggered by a finite initial phase velocity is not a mere classical crossover; it is fundamentally governed by the non-Markov state-dependent friction in Eq.~\eqref{langevin}, and can be interpreted as a dynamic topological phase transition associated with the onset of a zero-mode EP.  
To clarify its topological origin, we consider a response function 
\beq
G(t)=\left\langle \dfrac{\delta \phi_{0}(t)}{\delta f}\right\rangle ,
\label{response}
\eeq
where $f$ is an infinitesimal impulse force at $t=0$ \cite{SM}. 
Furthermore, we introduce a dynamic winding number around the origin $\omega=0$ in the complex frequency space \cite{volovik,gurarie2011}
\beq
\nu=\oint_{C}\dfrac{\dd{\omega}}{2\pi i}\dfrac{1}{\Tilde{G}(\omega)}\pdv{\Tilde{G}(\omega)}{\omega} ,
\label{windingnumber}
\eeq
where $C$ is a closed path along the semicircle enclosing all poles in the upper half plane. 
Figure \ref{Fignyquist}(a) shows the Fourier transform of the response function $\Tilde{G}(\omega) = \int_{0}^{t_{\rm rec}}\dd{t}G(t) e^{-i(\omega-i\eta)t}$ with $\eta>0$ being an infinitesimal positive constant for $\alpha=2$. 
The general form of the response function can be assumed as $\Tilde{G}(\omega)\sim [-(\omega-i\eta)^{2}-\Sigma(\omega)]^{-1}$ where $\Sigma(\omega)$ is the self-energy, including nonlinear contributions. 
As long as a finite friction is present, the self-energy has a positive imaginary term that involves a single pole at low energy \cite{SM}. 
When the non-Markov state-dependent friction perfectly cancels out at the critical velocity, the self-energy vanishes at low energy $\Sigma(\omega)\to0$. 
Consequently, the response function acquires a massless double pole exactly at the origin $\omega=0$ for $\eta\to0$, leading to the precise onset of the zero-mode EP. 
In the localized state with $v_{0}=10\Omega$ plotted by a solid curve, the response function in the frequency space has a finite winding around $\omega\simeq \pm10\Omega$, which corresponds to the oscillatory frequency of the Josephson phase. 
This topological property of the phase winding is rather clear in the complex frequency space.
As plotted in panel (b), $\Tilde{G}(\omega)$ ($\omega\in\mathbb{C}$) has single poles only in the lower half plane, which results in $\nu=0$, reflecting the translational symmetry breaking $\langle \cos{\phi_{0}}\rangle>0$ due to the Josephson mode trapped at the potential minimum. 
With a faster initial velocity, the state-dependent non-Markov dissipation averages the oscillatory frictional force yielding cancelation of the friction, and the Josephson mode turns into the running state as a free particle. 
As a result, once the Josephson potential becomes irrelevant, the Josephson mode effectively recovers the translational symmetry $\langle\cos{\phi_{0}}\rangle=0$. 
This recovery of translational symmetry due to the canceled friction indicates that the dynamic transition point is driven by the emergence of a zero-mode EP. 
The dashed curve in Fig.~\ref{Fignyquist}(a) shows that the running state with $v_{0}=20\Omega$ has the response function with a finite winding in the vicinity of $\omega=0$. 
In particular, the cardioid orbit indicates that the response function has a double pole at zero energy as a consequence of vanishing friction as well as the vanishing Josephson potential. 
In the complex frequency space, the response function possesses a double pole slightly shifted into the upper half plane resulting in $\nu=2$ \cite{SM}, which physically encapsulates the non-ergodic accelerating nature. 
Unlike conventional critical damping where poles merge at a finite decay rate in a damped harmonic oscillator, the highly nonlinear friction enforces the merging of the single poles exactly at zero energy. 
This specific onset of the zero-mode EP dictates the topological phase transition. 
This topological scenario is directly manifested in the real-time response function $G(t)$ [Fig.~\ref{Fignyquist}(c)]. 
In the localized phase near the critical point, $G(t)$ exhibits pronounced beating, a clear precursor of the merging poles. 
Notably, at the transition, this beating gives way to a persistent linear growth, a direct macroscopic hallmark of the anomalous algebraic response generated by the zero-mode EP. 
Note that the incoherent insulating state of the ground state in the SB phase diagram is characterized by $\nu=1$ because of finite friction as a consequence of Markov dissipation. 
The winding number \eqref{windingnumber} therefore plays the role of an order parameter for the localized-running dynamic phase transition that measures the zero mode and distinguishes the different types of incoherent phases induced by non-Markov dissipation from the conventional Markov one.

\begin{figure}[t]
\centering
\includegraphics[keepaspectratio,width=72mm]{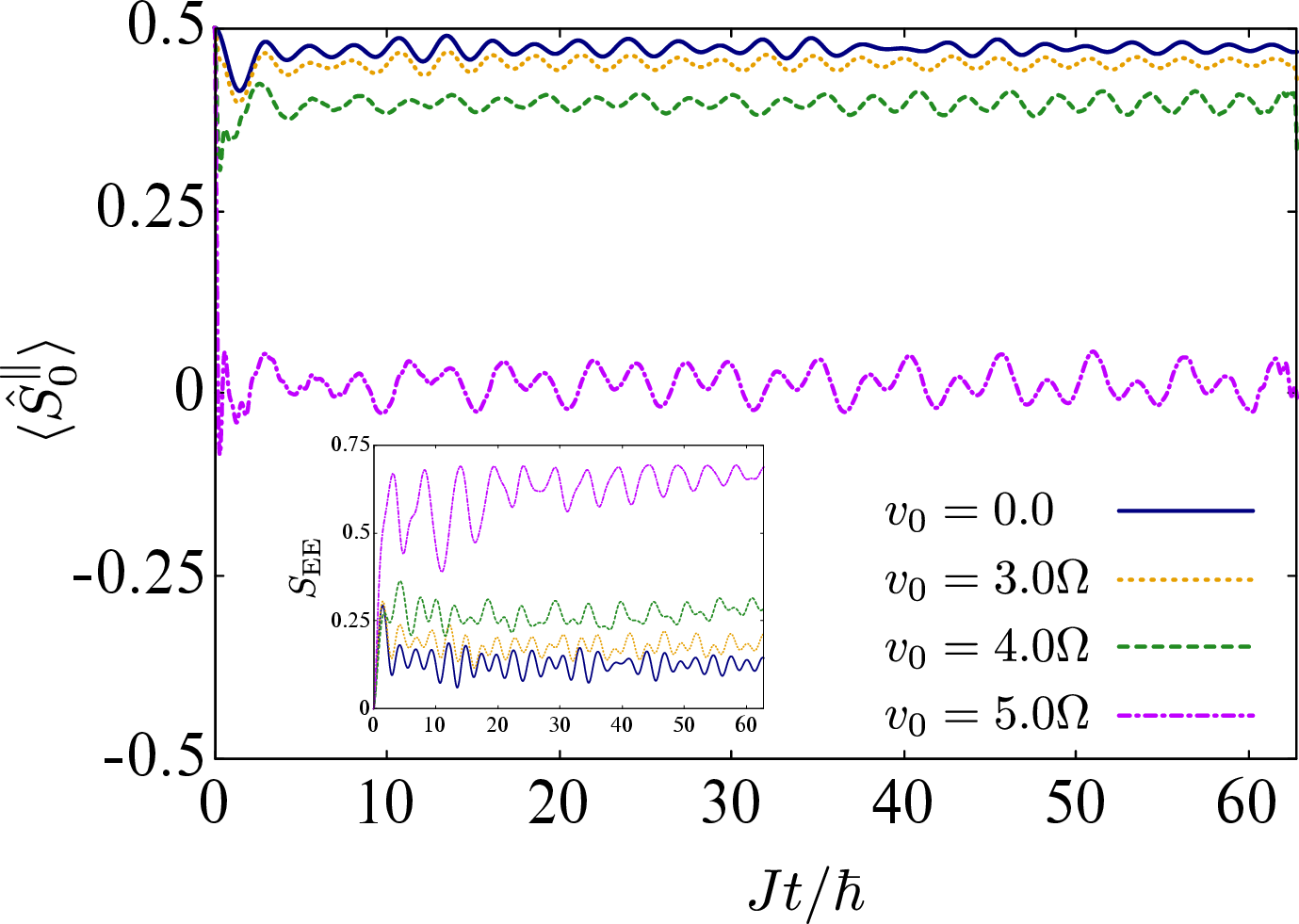}
\caption{Spin dynamics $\langle\hat{S}_{0}^{\parallel}(t)\rangle$ at the boundary in the driven XXZ spin chain \eqref{HdXXZ} obtained from the TEBD simulation for $\alpha=10$, $N=100$, and $\kappa L=10\pi$ with several values of $v_{0}$. 
Inset: Time evolution of the entanglement entropy $S_{\rm EE}$ between the boundary and the bulk XXZ spin chain. }
\label{FigSrot}
\end{figure}

{\it Robustness against quantum fluctuations and dynamical backreaction.---}
The non-Markov Langevin analysis incorporates both thermal and quantum fluctuations through Gaussian noise. While the non-Markov Langevin framework captures the essence of the topological transition, it treats the Josephson phase as a classical variable and enforces the fluctuation-dissipation relation. 
This inherently neglects quantum many-body backreaction from the out-of-equilibrium bath. 
To confirm the emergence of the running state induced by non-Markov dissipation even under strong quantum fluctuations and the dynamical backreaction, we show the TEBD simulation of the XXZ spin chain under a driving transverse magnetic field at the boundary, which is equivalent to the ICL model with a finite initial phase velocity at low energy. 
Through a unitary transformation mapping the ICL model to the bsG model \cite{furutani2024}, a finite initial velocity of the phase at the boundary in the ICL initial state leads to a multimode coherent state of the bath modes as the bsG initial state, which is an excited state. 
To demonstrate dynamics of the bsG model via the TEBD simulation of the XXZ spin chain, we perform a time-dependent unitary transformation to start with the vacuum ground state. 
The Hamiltonian after the unitary transformation reads \cite{SM}
\beq
\hat{H}_{\rm dbsG}=-E_{J}\cos{\qty[\dfrac{1}{\sqrt{\alpha}}\hat{\varphi}(0)+\Theta(t)]}+\hat{H}_{\rm TLL},
\label{HdbsG}
\eeq
where $\hat{H}_{\rm TLL}$ is the Tomonaga-Luttinger liquid (TLL) Hamiltonian and $\hat{\varphi}(x)$ is the phase operator in $\hat{H}_{\rm TLL}$. 
Here, the time-dependent driving field is given by
\beq
\Theta(t)=-\dfrac{2n_{0}^{3}}{\kappa L}\sqrt{\dfrac{2\pi}{\alpha}}\sum_{n=1}^{N}\dfrac{\Omega \sin{(\omega_{n}t)}}{\omega_{n}} ,
\label{Thetat}
\eeq
with $n_{0}\equiv \hbar v_{0}/2E_{C}$. 
Through bosonization \cite{tsvelik,giamarchi}, a driven XXZ model
\beq
\bal
\hat{H}_{\rm dXXZ}
&=J\sum_{j=1}^{N-1}\qty[\frac12 \left(\hat{S}_{j}^{+}\hat{S}_{j+1}^{-}+\hat{S}_{j}^{-}\hat{S}_{j+1}^{+}\right)+\Delta \hat{S}_{j}^{z}\hat{S}_{j+1}^{z}] \\
&+J'\qty[\frac12 \left(\hat{S}_{0}^{+}\hat{S}_{1}^{-}+\hat{S}_{0}^{-}\hat{S}_{1}^{+}\right)+\Delta' \hat{S}_{0}^{z}\hat{S}_{1}^{z}] \\
&-h\qty(e^{i\Theta(t)}\hat{S}_{0}^{+}+\text{h.c.}) ,
\eal
\label{HdXXZ}
\eeq
should give equivalent dynamics as the driven bsG model \eqref{HdbsG} at low energy. 
Figure \ref{FigSrot} shows the dynamics of $\langle\hat{S}_{0}^{\parallel}(t)\rangle=\langle\hat{S}_{0}^{x}(t)\rangle\cos{\Theta(t)}-\langle\hat{S}_{0}^{y}(t)\rangle\sin{\Theta(t)}$, which is the component parallel to the magnetic field that plays the role of the coherence factor in this frame, obtained through the TEBD simulation in the deep coherent regime $\alpha=10$. 
Initially, we set $J'=0$ and the bulk XXZ spin chain is in the ground state. 
The boundary spin $\hat{\bm{S}}_{0}$ is initially pinned by a giant magnetic field and $\langle\hat{S}^{x}_{0}(0)\rangle=0.5$. 
Then, we turn on the coupling $(J',\Delta')=(J,\Delta)$ at $t>0$. 
In this deep coherent regime, as $v_{0}$ increases, the time average of the parallel spin component $\langle\hat{S}_{0}^{\parallel}\rangle$ gradually decreases and drops for $v_{0}\simeq 5.0\Omega$, which is close to the critical velocity predicted from the above Langevin analysis. 
The inset presenting the entanglement entropy $S_{\rm EE}$ between the boundary and the bulk XXZ spin chain indicates strong entanglement by increasing $v_{0}$. 
Crucially, the simultaneous growth of the entanglement entropy $S_{\rm EE}$ indicates a severe breakdown of the Born-Markov approximation, proving that the bath modes are strongly driven out of equilibrium. 
Yet, the macroscopic running behavior persists. 
These exact TEBD results definitively demonstrate that the topological signature of the zero-mode EP robustly governs the macroscopic fate, surviving both strong quantum fluctuations and the dynamical backreaction from the environment.

%\section{Conclusion}
To conclude, we have demonstrated a fundamentally new class of dynamical phase transitions in the intrinsically momentum-coupled Caldeira-Leggett model, microscopically derived from the head-to-tail Bose-Josephson junction. 
Unlike linear or conventional Markov open quantum systems that inevitably thermalize to their ground state, our exact nonlinear non-Markov Langevin dynamics reveals a striking breakdown of ergodicity. 
The macroscopic phase dynamics entirely evades thermalization, governed instead by a sharp transition into an incoherent running state. 
Crucially, we established that this non-ergodic transition is topologically driven by a zero-mode exceptional point. 
The merging of poles of the response function, induced by highly nonlinear non-Markov friction, generates a massless double pole that is rigorously characterized by a dynamic winding number.

Although the resulting running state is phenomenologically similar to macroscopic quantum self-trapping (MQST) \cite{shenoy1999, albiez2005, bardin2024, bardin2026}, their physical origins are profoundly distinct. 
Conventional MQST, the onset of the pinned population imbalance throughout time evolution, originates from a standard pitchfork bifurcation in the mean-field energy landscape under conservative dynamics. 
In sharp contrast, our running phase is a topological consequence of non-Markov dissipation, a mechanism entirely forbidden in the Markov limit. 
Furthermore, our TEBD simulations of the equivalent driven XXZ spin chain confirm that this exceptional-point-induced signature survives as a robust dynamical crossover, resisting not only strong quantum and thermal fluctuations, but also the dynamical backreaction from out-of-equilibrium distribution of the bath modes.

Our findings challenge the standard ergodic picture in structured environments, revealing that initial conditions can deterministically dictate macroscopic long-time fates via non-Markovian topology. 
This nontrivial interplay between nonlinearity and structured dissipation redefines the role of environment in open quantum systems. 
Our theoretical insights open new experimental frontiers in highly controllable ultracold atomic gasses and superconducting circuits, offering a novel paradigm for exploring interaction-driven non-Markov dynamics and protecting long-lived quantum memory.

\begin{acknowledgements}
The author thanks Yuki Kawaguchi for fruitful discussions. 
K.F. was supported by JSPS KAKENHI (Grant No. JP24K22858, No. JP24K00557, and No. JP26H00385) and Murata Science and Education Foundation. 
\end{acknowledgements}

\bibliography{reference.bib}

\newpage
\appendix
%%%%%%%%%%%%%%%% Supplemental Material %%%%%%%%%%%%%%
\widetext
\pagebreak
\begin{center}
\textbf{\large Supplemental Material for "Non-ergodic dynamical phase transition via a zero-mode exceptional point in a non-Markov atomic Josephson junction"}
\end{center}

\renewcommand{\theequation}{S\arabic{equation}}
\renewcommand{\thefigure}{S\arabic{figure}}
\renewcommand{\bibnumfmt}[1]{[S#1]}
\setcounter{equation}{0}
\setcounter{figure}{0}

\section{Markov embedding of non-Markov Langevin equation}\label{Appembedding}

In this section, we provide the details of the numerical calculation of Eq.~(2).  
To this end, we introduce a complex dynamical variable 
\beq
Z_{n}(t)\equiv \int_{0}^{t}\dd{s} e^{i\omega_{n}s}\cos{\phi_{0}(s)} \dot{\phi}_{0}(s). 
\eeq
Using the damping kernel in Eq.~(3), one can decompose the non-Markov Langevin equation (2) into a set of $(N+2)$ Markov equations as
\bseq
\beq
\dot{\phi}_{0}(t)=v(t),
\label{phidot}
\eeq
\beq
\bal
\dot{v}(t)&=-\Omega^{2}\displaystyle\sum_{n=1}^{N}\Big[\cos{(\omega_{n}t)} \mathrm{Re}[Z_{n}(t)] + \sin{(\omega_{n}t)} \mathrm{Im}[Z_{n}(t)]\Big] 
-\Omega^{2}\sin{\phi_{0}(t)} +\dfrac{1+N}{E_{J}} \Omega^{2}\xi(t), 
\eal
\label{vdot}
\eeq
\beq
\dot{Z}_{n}(t)=e^{i\omega_{n}t} \cos{\phi_{0}(t)} v(t) .
\label{Zdot}
\eeq
\label{markovianeq}
\eseq
The microscopic definition of the stochastic noise $\xi(t)$ is given by
\beq
\xi(t)=-\dfrac{\hbar^2}{2E_{C} L}\sum_{n=1}^{N}\qty[\omega_{n}^{2}\cos{(\omega_{n}t)}Q_{n}(0)+\dfrac{\omega_{n}}{M}\sin{(\omega_{n}t)}P_{n}(0)],
\label{xiQPrep}
\eeq
where $(Q_{n},P_{n})$ is the coordinate and the momentum of the $n$th bath mode. 
The phonon bath is initially assumed to be in thermal equilibrium at temperature $T$. 
That allows us to introduce a stochastic variable $a_{n} \in\mathbb{C}$ defined by
\beq
Q_{n}(0)=\sqrt{\dfrac{\hbar}{2M\omega_{n}}} \qty(a_{n}+a_{n}^{*}), \quad\quad
P_{n}(0)=-i\sqrt{\dfrac{M\hbar\omega_{n}}{2}} \qty(a_{n}-a_{n}^{*}) ,
\eeq
satisfying
\beq
\langle a_{n}\rangle = \langle a_{n}^{*}\rangle=0, \quad\quad
\langle a_{n}^{*}a_{m}\rangle=\langle a_{m}^{*}a_{n}\rangle
=\frac12 \coth{\qty(\dfrac{\hbar\omega_{n}}{k_{\rm B}T})}\delta_{n,m},
\eeq
with $\langle\cdots\rangle$ being the thermal average. 
By using the stochastic variables, the original noise \eqref{xiQPrep} can be written as
\beq
\xi(t)=-\dfrac{\hbar^2}{2E_{C} L}\sum_{n=1}^{N}\sqrt{\dfrac{\hbar\omega_{n}^{3}}{2M}}\qty(a_{n} e^{-i\omega_{n}t}+a_{n}^{*} e^{i\omega_{n}t}),
\eeq
which meets Eqs.~(5). 
Then, one can write Eq.~\eqref{vdot} as
\beq
\bal
\dot{v}(t)&=-\Omega^{2}\displaystyle\sum_{n=1}^{N}\Big[\cos{(\omega_{n}t)} \mathrm{Re}[\Tilde{Z}_{n}(t)] + \sin{(\omega_{n}t)} \mathrm{Im}[\Tilde{Z}_{n}(t)]\Big] 
-\Omega^{2}\sin{\phi_{0}(t)},
\eal
\label{vdotan}
\eeq
with 
\beq
\Tilde{Z}_{n}(t)\equiv Z_{n}(t)+2 \qty(\dfrac{\omega_{n}}{\Omega})^{3/2}\sqrt{\dfrac{2\pi}{\alpha \kappa L}} \, a_{n}
\eeq
We solved Eqs.~\eqref{phidot}, \eqref{vdotan}, and \eqref{Zdot} under a given initial condition $(\phi_{0}(0), \dot{\phi}_{0}(0)=v_{0}, Z_{n}(0)=0)$. 
In numerically computing the response function (6), %\eqref{response}, 
we prepared two orbits with two different initial conditions $(\phi_{0}(0), v_{0},Z_{n}(0)=0)$ and $(\phi_{0}(0), v_{0}+\Delta v_{0},Z_{n}(0)=0)$ with $\Delta v_{0}=10^{-7}\Omega$ being an infinitesimal impulse, and subtracted the two orbits.

\section{Response functions}\label{Appcardioid}

In the Markov limit under the linear approximation, the response function has the form of 
\beq
\Tilde{G}_{\rm linear}(\omega)\propto\dfrac{1}{-\omega^{2}-2i\gamma_{0}\omega +\Omega^{2}},
\label{Gmarkov}
\eeq
where $\gamma_{0}$ stands for the damping constant in the Markov limit. 
In the non-Markov and nonlinear regime, however, it is highly nontrivial to obtain the response function. 
To verify that the nonlinear non-Markov friction in Eq.~(2) is washed out, let us assume a response function in the following form
\beq
\Tilde{G}(\omega)=\dfrac{Z^{-1}(\omega)}{-(\omega-i\eta)^{2}-\Sigma(\omega)},
\eeq
where $Z(\omega)$ represents the correction to the mass term by the nonlinear contributions and $\Sigma(\omega)$ is the self-energy. 
The irrelevance of friction and the Josephson potential corresponds to the vanishing self-energy $\Sigma(\omega)=0$. 
As long as $\abs{Z^{-1}(\omega)}$ is constant, the modulus of the response function then reads
\beq
\abs{\Tilde{G}(\omega)}=\frac{\abs{Z^{-1}}}{2\eta^{2}} \qty[1-\cos{(2\varphi)}],
\label{cardioid}
\eeq
with $\varphi=\arg{[\omega-i\eta]}$. 
Equation~\eqref{cardioid} represents the cardioid curve. 
In other words, a cardioid orbit of $\Tilde{G}(\omega)$ at $\omega=0$ indicates a double pole as $\Tilde{G}(\omega)\propto (\omega-i\eta)^{-2}$. 

The incoherent insulating state in the Schmid-Bulgadaev equilibrium phase diagram in the region of $\alpha<1$ has distinct singularities. 
Indeed, the friction term in Eq.~\eqref{Gmarkov} survives in the Markov limit and the kinetic term $\propto\omega^{2}$ is no longer relevant at low energy \cite{altlandsimons}, which fully reproduces the Schmid-Bulgadaev phase diagram exactly obtained by solving the integrable boundary sine-Gordon model. 
The negligible kinetic term indicates that the response function \eqref{Gmarkov} can have only a single pole in contrast to the non-Markov case.

\section{Driven boundary sine-Gordon model}

The generalized Langevin equation (2) results from the ICL Hamiltonian
\beq
\hat{H}_{\rm ICL}=E_{C}\hat{N}^{2}-E_{J}\cos{\hat{\phi}_{0}}+\sum_{n=1}^{N}\hbar\omega_{n}\hat{b}_{n}^{\dagger}\hat{b}_{n}
-\hat{N}\sum_{n=1}^{N}\hbar\kappa_{n}(\hat{b}_{n}^{\dagger}+\hat{b}_{n}),
\label{HICL}
\eeq
with
\beq
E_{C}\equiv \dfrac{(1+N)\hbar^{2}}{2ML^{2}}, \quad\quad
E_{J}\equiv J_{0}L\bar{n}, \quad\quad
\hbar\kappa_{n} \equiv \sqrt{\dfrac{E_{C}\hbar\omega_{n}}{1+N}}.
\eeq
The ICL model can be mapped to the bsG model through a unitary transformation by $\hat{U}\equiv \mathrm{exp}[-i\hat{N}\hat{\Xi}]$ with $\hat{\Xi}\equiv i\sum_{n=1}^{N}\kappa_{n}(\hat{b}_{n}^{\dagger}-\hat{b}_{n})/\omega_{n}$ as \cite{ashida2022,masuki2022,furutani2024}
\beq
\bal
\hat{H}_{\rm bsG} &=\hat{U}^{\dagger}\hat{H}_{\rm ICL}\hat{U} 
=-E_{J}\cos{\qty[\hat{\phi}_{0}+\dfrac{1}{\sqrt{\alpha}}\hat{\varphi}(0)]}
+\hat{H}_{\rm TLL}, \quad\quad
\hat{H}_{\rm TLL}=\dfrac{\hbar c}{4\pi}\int_{0}^{L}\dd{x}\qty[(\partial_{x}\hat{\varphi})^{2}+\hat{\Pi}^{2}] .
\eal
\eeq
From the ICL Hamiltonian \eqref{HICL}, the equation of motion gives
\beq
\dv{\hat{\phi}_{0}}{t}=\dfrac{1}{i\hbar}[\hat{\phi}_{0},\hat{H}_{\rm ICL}]
=\dfrac{2E_{C}}{\hbar}\hat{N} - \sum_{n=1}^{N}\kappa_{n}(\hat{b}_{n}^{\dagger}+\hat{b}_{n}),
\label{dphidt}
\eeq
where we used the canonical commutation relation $[\hat{\phi}_{0},\hat{N}]=i$. 
Our Langevin simulation in the main text starts from an initial state with a finite Josephson phase velocity $v_{0}=\langle\partial_{t}\hat{\phi}_{0}(t=0)\rangle$. 
Equation \eqref{dphidt} then gives a relation
\beq
v_{0}=\dfrac{2E_{C}}{\hbar}\langle\hat{N}(t=0)\rangle \equiv \dfrac{2E_{C}n_{0}}{\hbar}.
\eeq
Using $n_{0}\equiv \langle\hat{N}(t=0)\rangle$, the initial state can be expressed as a direct product of the Josephson mode and the bath mode as
\beq
\ket{\rm ICL}_{0} \equiv \ket{n_{0}}_{\rm J}\otimes \ket{0}_{\rm B},
\eeq
where $\ket{n_{0}}_{\rm J}$ represents the eigenstate of $\hat{N}$ with the eigenvalue $n_{0}$ and $\ket{0}_{\rm B}$ is the vacuum of the phonon bath. 
The corresponding initial state of the bsG model reads
\beq
\bal
\ket{\rm bsG}_{0} &\equiv \hat{U}^{\dagger}\ket{\rm ICL}_{0} \\
&=e^{i\hat{N}\hat{\Xi}} \ket{n_{0}}_{\rm J}\otimes \ket{0}_{\rm B} 
=\ket{n_{0}}_{\rm J}\otimes e^{in_{0}\hat{\Xi}}\ket{0}_{\rm B} 
=\ket{n_{0}}_{\rm J}\otimes \prod_{n=1}^{N}\hat{D}(\alpha_{n})\ket{0}_{\rm B},
\eal
\label{bsGinitial}
\eeq
where $\hat{D}(\alpha_{n})\equiv \mathrm{exp}[\alpha_{n}\hat{b}_{n}^{\dagger}-\alpha_{n}^{*}\hat{b}_{n}]$ is the displacement operator with 
\beq
\alpha_{n}\equiv -\dfrac{n_{0}}{L}\sqrt{\dfrac{\hbar}{2M\omega_{n}}} .
\eeq
As a result, it turned out that the initial state of the bsG model is a direct product of the Josephson mode and the multiple coherent state of the bath modes $\ket{\rm bsG}_{0}=\ket{n_{0}}_{\rm J} \otimes \ket{\alpha_{1},\alpha_{2},\cdots,\alpha_{N}}_{\rm B}$. 
When we start with a sufficiently large phase velocity, the density $n_{0}$ becomes large as well indicating high occupancy of each phonon in the bath. 
For numerical convenience, we consider to find an appropriate frame with effectively low occupancy of the bath modes. 
To this end, we first introduce the interaction representation of $\hat{\Xi}$ as
\beq
\hat{\Xi}(t)\equiv -i\sum_{n=1}^{N}\qty[\alpha_{n}(t)\hat{b}_{n}^{\dagger}-\alpha_{n}^{*}(t)\hat{b}_{n}], \quad\quad
\alpha_{n}(t)\equiv \alpha_{n} e^{i\omega_{n}t},
\label{Xit}
\eeq
which satisfies 
\beq
\hat{\Xi}(0)=\hat{\Xi}, \quad\quad 
\dv{\hat{\Xi}(t)}{t}=\sum_{n=1}^{N}\omega_{n}\qty[\alpha_{n}(t)\hat{b}_{n}^{\dagger}+\alpha_{n}^{*}(t)\hat{b}_{n}] .
\eeq
With Eq.~\eqref{Xit}, we introduce a time-dependent unitary transformation given by
\beq
\hat{V}(t)\equiv e^{in_{0}[\hat{\phi}_{0}+\hat{\Xi}(t)]} .
\eeq
The time-dependent unitary transformation maps the initial state \eqref{bsGinitial} to
\beq
\ket{\rm dbsG}_{0}=\hat{V}^{\dagger}(0)\ket{\rm bsG}_{0}
=e^{-in_{0}(\hat{\phi}_{0}+\hat{\Xi})}\ket{n_{0}}_{\rm J}\otimes e^{in_{0}\hat{\Xi}}\ket{0}_{\rm B}
=\ket{0}_{\rm J}\otimes \ket{0}_{\rm B}.
\label{dbsGinitial}
\eeq
In the last equality, we used the relation
\beq
\hat{N}e^{-in_{0}\hat{\phi}_{0}}\ket{n_{0}}_{\rm J}
=\qty(e^{-in_{0}\hat{\phi}_{0}}\hat{N}+[\hat{N},-in_{0}\hat{\phi}_{0}]e^{-in_{0}\hat{\phi}_{0}})\ket{n_{0}}_{\rm J}
=e^{-in_{0}\hat{\phi}_{0}}(\hat{N}-n_{0})\ket{n_{0}}_{\rm J}
=0,
\eeq
which leads to $\ket{0}_{\rm J}=\mathrm{exp}[-in_{0}\hat{\phi}_{0}] \ket{n_{0}}_{\rm J}$. 
Equation \eqref{dbsGinitial} indicates that the transformed initial state is vacuum. 
The transformed Hamiltonian is given by
\beq
\bal
\hat{H}_{\rm dbsG}(t) &=\hat{V}^{\dagger}(t)\hat{H}_{\rm bsG}\hat{V}(t)-i\hbar\dv{\hat{V}^{\dagger}(t)}{t}\hat{V}(t) \\
&=-E_{J}\cos{\qty[\hat{\phi}_{0}+\dfrac{1}{\sqrt{\alpha}}\hat{\varphi}(0)+\Theta(t)]}+\hat{H}_{\rm TLL} ,
\eal
\label{HdbsG1}
\eeq
where $\Theta(t)$ is the time-dependent driving field at the boundary given by Eq.~(9). 

Through bosonization \cite{tsvelik,giamarchi}, the driven bsG model \eqref{HdbsG1} can be obtained from a XXZ spin chain with a time-dependent transverse magnetic field described by Eq.~(10). 
One can therefore numerically verify the onset of the non-Markov-dissipation-induced running phase transition by the time evolution of the boundary-driven quantum spin chain from the ground state.

\section{Relaxation to the ground state}

\begin{figure}[t]
\centering
\includegraphics[keepaspectratio,width=72mm]{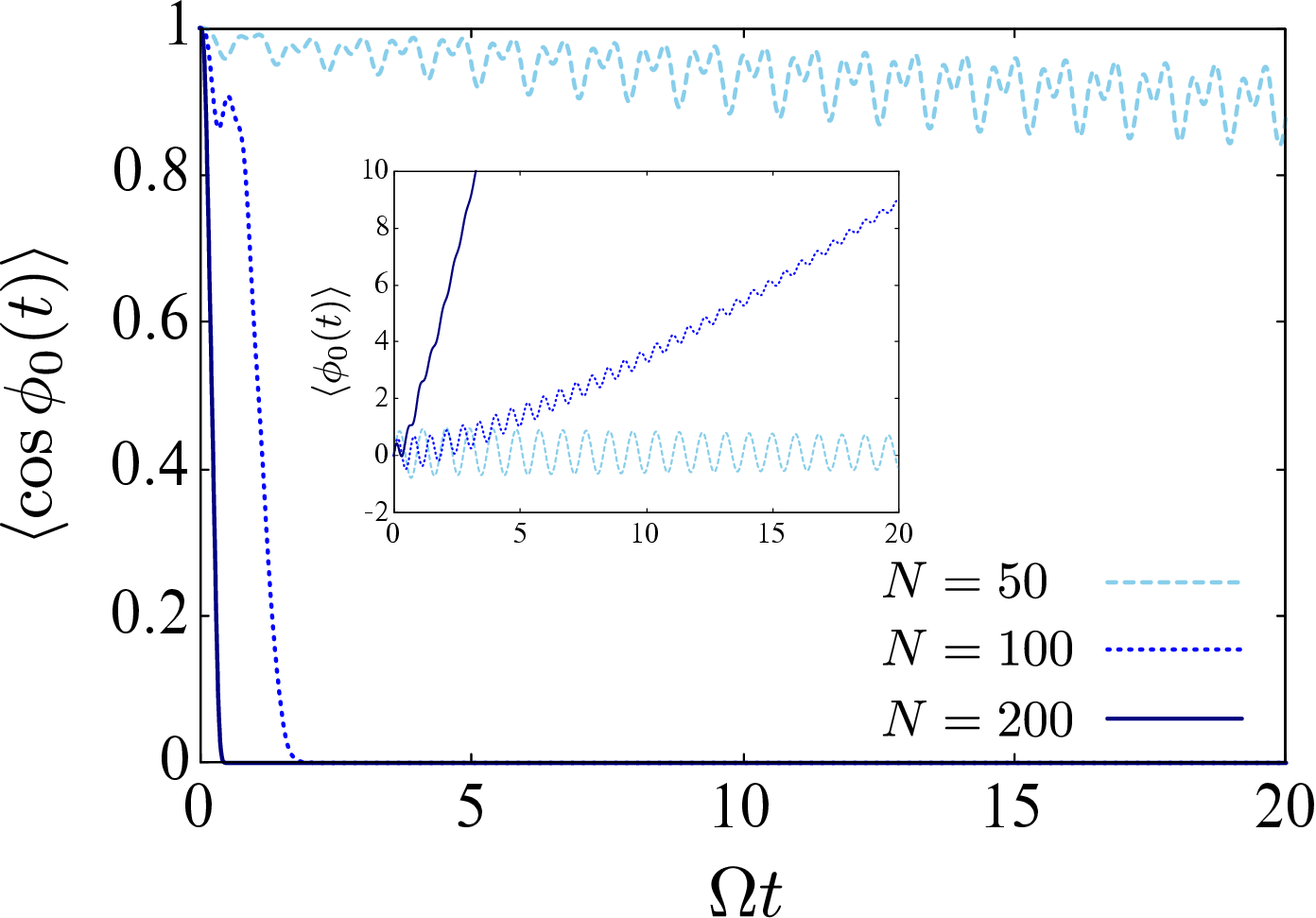}
\includegraphics[keepaspectratio,width=72mm]{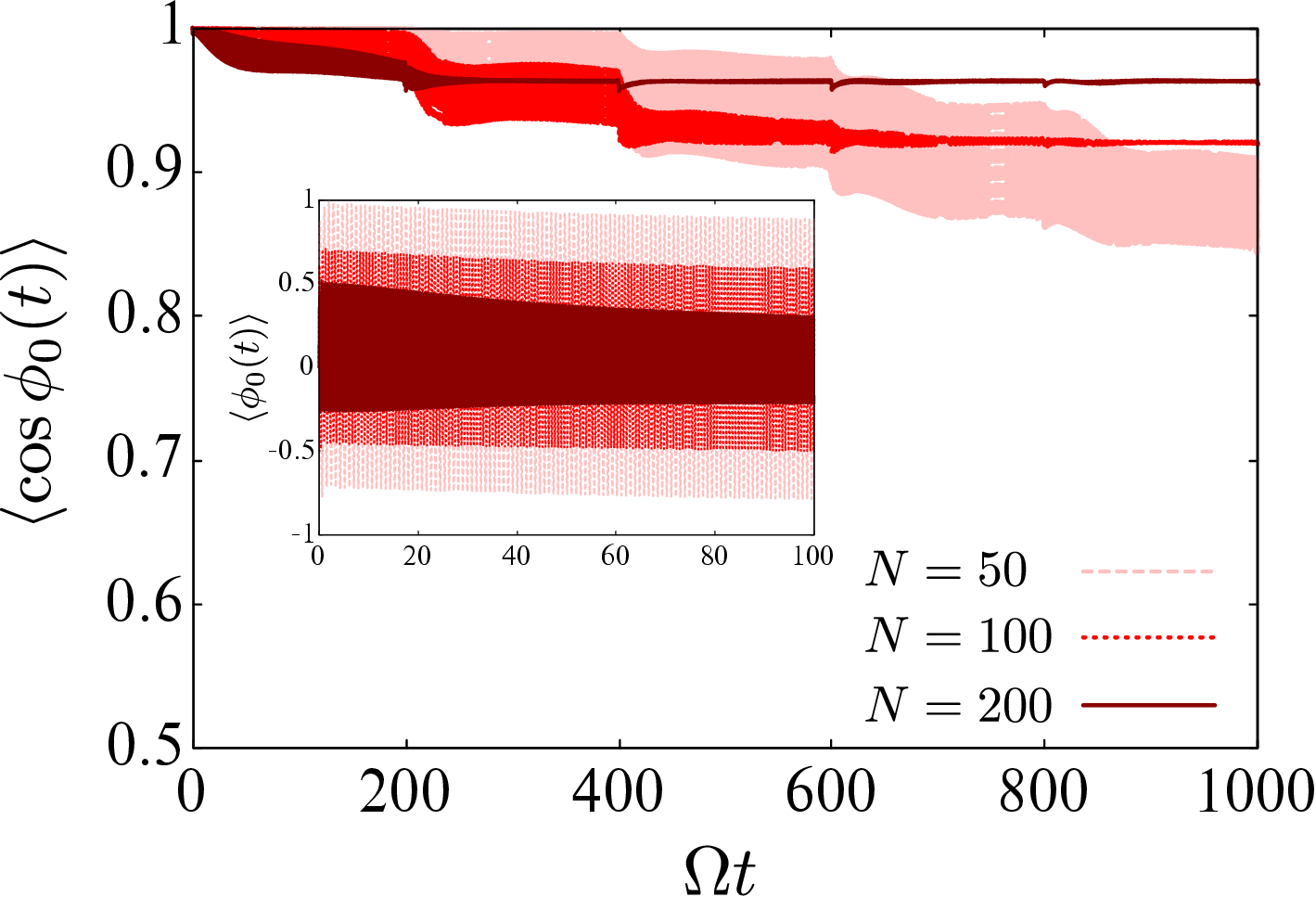}
\caption{Time evolution of the coherence factor $\langle\cos{\phi_{0}(t)}\rangle$ for $\alpha=0.1$ (left) and $\alpha=10$ (right), respectively, under $\kappa L = 10^2$, $T=0$, and $v_{0}=6\Omega$. 
Inset: Time evolution of the Josephson phase $\langle\phi_{0}(t)\rangle$. }
\label{FigMarkov}
\end{figure}

In the Markov limit $N\to\infty$, the ground state of the BJJ is known to be superconducting for $\alpha>1$ while insulating for $\alpha<1$ \cite{furutani2024}. 
We can confirm that the BJJ relaxes to the ground state in $N\to\infty$ according to Fig.~\ref{FigMarkov}, which shows the coherence factor $\langle\cos{\phi_{0}(t)}\rangle$. 
For $\alpha=0.1$ in the left panel, it approaches zero as the number of modes $N$ increases indicating the incoherent insulating state. 
Indeed, the phase average in the inset shows a transition into running phase from the Josephson oscillations. 
For $\alpha=10$ in the right panel, on the other hand, the coherence tends to remain unity by increasing $N$, as also confirmed by the dwelling amplitude of the Josephson oscillations of the phase average, indicating the coherent superconducting state.

\section{Running phase transition at finite temperatures}

\begin{figure}[t]
\centering
\includegraphics[keepaspectratio,width=72mm]{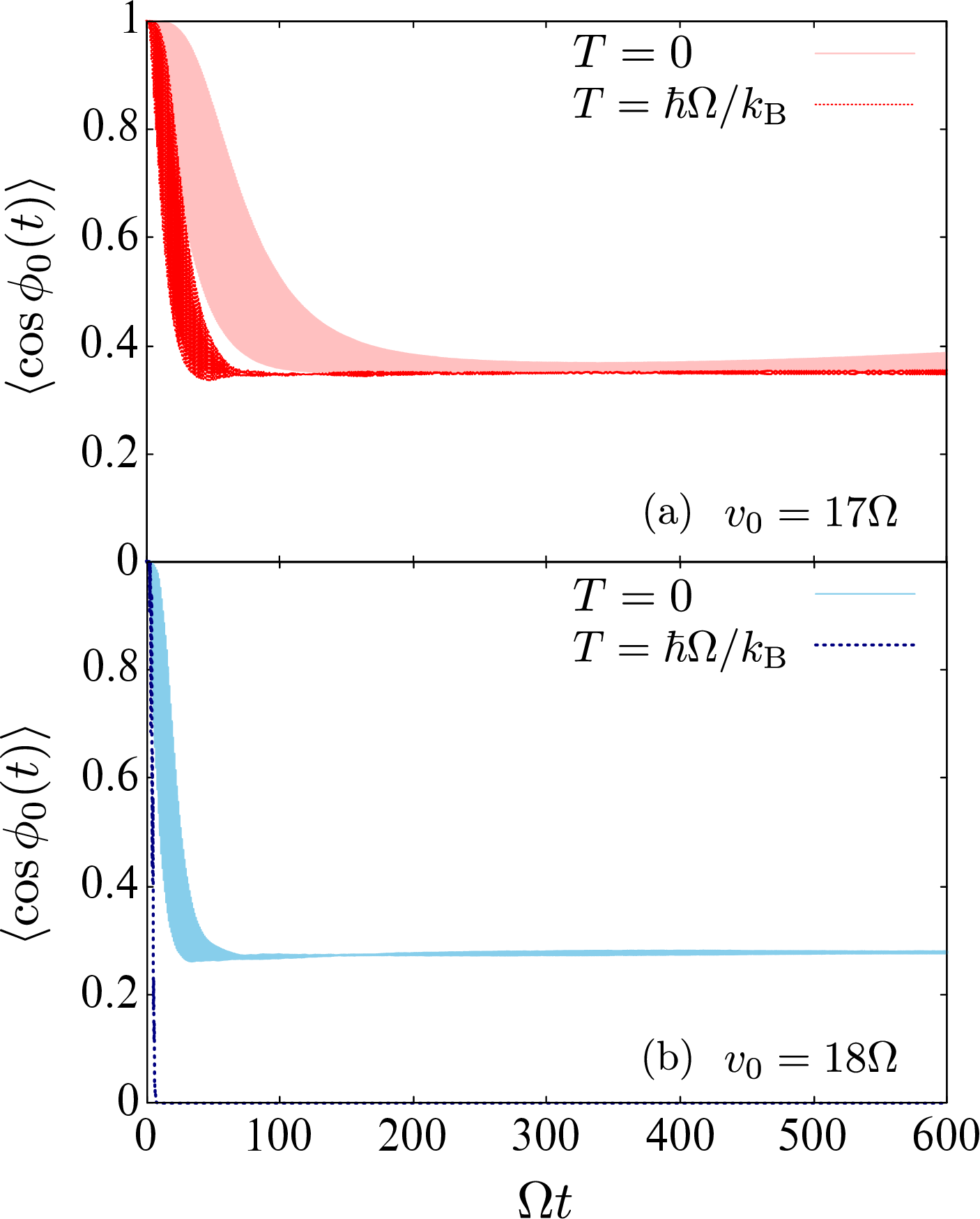}
\caption{Time evolution of the coherence factor $\langle\cos{\phi_{0}(t)}\rangle$ for $T=0$ and $T=\hbar\Omega/k_{\rm B}$, respectively, under $N=100$ and $\kappa L = 10^2$ with different initial velocities: (a) $v_{0}=17\Omega$, (b) $v_{0}=18\Omega$. }
\label{FigcoherencefiniteT}
\end{figure}

Thermal Gaussian noise incorporates the effect of thermal fluctuations. 
In this section, we show that the running phase transition predicted at zero temperature is robust against thermal fluctuations. 
Figure \ref{FigcoherencefiniteT} plots the coherence factor for the different initial velocities at finite temperature. 
In the localized phase with $v_{0}=17\Omega$, the coherence factor remains finite at finite temperature $T=\hbar\Omega/k_{\rm B}$. 
In the vicinity of the critical velocity at zero temperature with $v_{0}=18\Omega\lesssim v_{0,\mathrm{c}}$, it remains finite at zero temperature, but it sharply drops at $T=\hbar\Omega/k_{\rm B}$. 
It tells us that the emergence of the non-Markov dissipation-induced dynamic phase transition remains even at finite temperature, while thermal fluctuations quantitatively decrease the critical velocity, preferring the incoherent delocalized state.

\end{document}